\documentclass[11pt]{article}
\usepackage{amsfonts}
\usepackage{amsmath}
\usepackage{geometry}
\usepackage{graphicx}
\usepackage{amssymb}

\input{tcilatex}
\begin{document}

\title{A mathematical model of kinetoplastid mitochondrial gene scrambling
advantage}
\author{H. Buhrman\thanks{%
Centrum Wiskunde \& Informatica, and University of Amsterdam} \and P. T. S.
van der Gulik\thanks{%
Centrum Wiskunde \& Informatica} \and S. Severini\thanks{%
University College London} \and D. Speijer\thanks{%
Corresponding author: Department of Medical Biochemistry, Academic Medical
Center, University of Amsterdam, Meibergdreef 15, 1105 AZ Amsterdam, The
Netherlands; E-mail: D.Speijer@amc.uva.nl}}
\maketitle

\begin{abstract}
We model and discuss advantages of pan-editing, the complex way of
expressing mitochondrial genes in kinetoplastids. The rapid spread and
preservation of pan-editing seems to be due to its concomitant fragment
dispersal. Such dispersal prevents losing temporarily non expressed
mitochondrial genes upon intense intraspecific competition, by linking non
expressed fragments to parts which are still needed. We mathematically
modelled protection against gene loss, due to the absence of selection, by
this kind of fragment association. This gives ranges of values for
parameters like scrambling extent, population size, and number of
generations still retaining full genomes despite limited selection. Values
obtained seem consistent with those observed. We find a quasi-linear
correlation between dispersal and number of generations after which
populations lose genes, showing that pan-editing can be selected to
effectively limit gene loss under relaxed selective pressure.
\end{abstract}

\section{Introduction}

Gene fragmentation and dispersal of fragments are found in various organisms
ranging from Euglenozoa (found in kinetoplastids, euglenoids, and
diplonemids) to Alveolates (found in apicomplexa, ciliates, and
dinoflagellates). See (Benne \emph{et al.}, 1986; Gillespie \emph{et al.},
1999; Marande and Burger, 2007; Kamikawa \emph{et al.}, 2007; Nowacki \emph{%
et al.}, 2008; Walker, 2007; Spencer and Gray, 2011). Gene fragmentation
occurs when the RNA is not made from a single precursor, derived directly
from the genome, but it is reconstituted from different RNA sources. This
can also happen without dispersal, \emph{i.e. }spreading the genes for these
RNA parts over the genome, intermingled with other genes.

RNA editing, first used to refer to RNA alteration processes in
trypanosomatids, entailing extreme mitochondrial (mt) gene fragment
dissemination, was discovered in 1986 (Benne \emph{et al.}, 1986).
Trypanosomatids are parasitic unicellular organisms belonging to the
kinetoplastid order. Kinetoplastids are characterized by a 'kinetoplast',
the strongly staining location of \emph{all} the mt DNA molecules within
their \emph{single} mitochondrion. Kinetoplastid RNA editing refers to
post-transcriptional sequence alteration via insertion and deletion of
uridylate residues at specific sites of mt RNAs.

Trypanosomal mt RNAs are encoded by two types of mt DNA molecules: several
thousand small minicircles and a few dozen maxicircles (Scott, 1995; Simpson
and Thiemann, 1995; Simpson \emph{et al.}, 2000; Madison-Antenucci \emph{et
al.}, 2002). Maxicircles contain ribosomal RNA genes and genes mostly
encoding subunits of respiratory chain complexes. Many subunit genes need
editing upon transcription. Both kind of circles encode guide (g) RNAs
necessary for editing of maxicircle encoded RNAs (Blum \emph{et al.}, 1990).

Very extensive editing is illustrated by the \emph{Trypanosoma brucei} cox3
gene. Cox3 is easily found (\emph{e.g.} by looking at homologies) in \emph{%
Crithidia fasciculata} and \emph{Leishmania tarentolae}: the transcript
undergoes limited editing (\emph{e.g.} in \emph{C. fasciculata} only 32 Us
are inserted and 2 Us deleted at 14 sites).

The \emph{T. brucei} cox3 'gene' was identified later: its transcript has to
be edited overall (Feagin \emph{et al.}, 1988). While the G, A and C
nucleotide sequence is maxicircle derived, the U nucleotide sequence is
generated by editing, using minicircle encoded gRNAs, with $547$ Us inserted
and $41$ Us deleted at $223$ sites. The transcript derived from the 'GAC
sequence' (called the 'cryptogene' (Simpson and Shaw, 1989)) has less than
half the edited size.

In \emph{T. brucei}, most of the mt encoded proteins come from such
cryptogenes ($9$ out of $17$, with $5$ unedited and $3$ partially edited
transcripts remaining). The extensive editing of cryptogene transcripts is
known as pan-editing (Simpson and Shaw, 1989; Sturm and Simpson, 1990). The
precise molecular mechanisms of editing, using sequential basepairing
between gRNAs and mRNA, are described in (Schnaufer \emph{et al.}, 2003) and
references therein (Madison-Antenucci \emph{et al.}, 2002; Blum \emph{et al.}%
, 1990; Feagin \emph{et al.}, 1988; Simpson and Shaw, 1989). The consequence
of pan-editing is that information necessary for production of (some)
proteins is spread and mixed over the entire mt DNA.

Here, we propose a mathematical model to support the concept (Speijer, 2006;
Speijer, 2007) that such gene fragment dispersal can function as a
biological adaptation preventing gene loss caused by intense intraspecific
competition, due to selection on growth rate in limited spaces (\emph{e.g.}
a host). The process lowers the chance that rapidly growing deletion mutants
replace \emph{all} the more complex, ecologically versatile, organisms, as
large deletions will contain linked 'active' segments. It thus works to
'preserve by association'. To put it another way: mt gene scattering
prevents the translation of short-term advantages into long-term disaster.

\section{Modelling the effects of mitochondrial gene fragment dispersal in
changing environments}

Our model describes the `life cycle' of the mt genetic material of a
trypanosomatid. In the absence of functional selection we do not distinguish
between competition at the levels of individual mt genomes (\emph{e.g.}
maxicircles) and whole organisms. At the molecular level small circles
outcompete larger circles by faster replication while at the level of the
organism, organisms carrying less mt DNA can also grow faster. \emph{At both
levels mt deletions will be selected then: they even reinforce each other.}
We need to include the following parameters to determine the relationship
between some of them:

\begin{enumerate}
\item Population size ($s_{p}$), $10^{10}$ individuals;

\item Deletion size ($l$);

\item Probability of a deletion ($p$), $10^{-6}$;

\item Level of fragmentation/editing ($k$);

\item Number of generations in each environment (in the case of parasites
`host');

\item Replication advantage of smaller DNA ($r(x)$);

\item Confidence parameter ($d$) that governs the level of approximation in
our second simulation.
\end{enumerate}

As mitochondrial DNA polymerases are much more error-prone than their
nuclear counterparts (see (Larsson, 2010) and references therein), a value
of $10^{-6}$ for the deletion probability is not unrealistic. We first model
the process of replication and deletion in its most general form, and then
simplify it, to enable us to simulate for realistic values of our
parameters. In all our simulations, we assume there are two environments, $A$
and $B$, \emph{e.g.} $A$, tsetse fly and $B$, human host. A sequence '$a$'
of base pairs of length $n_{a}$ is necessary for survival in $A$ and a
sequence '$b$' of length $n_{b}$ is needed in $B$, but not in $A$. For
simplicity we excluded sequences needed in both (violet segments in Figure
1). Total length of mt DNA will therefore be $n=n_{a}+n_{b}$ base pairs. We
chose $n_{a}$ to represent $65\%$ and $n_{b}$ $35\%$ of total mt DNA ($\sim
23000$ basepairs). We assume the parasite to be in environment $A$.

The level of mt DNA fragmentation for an individual parasite is simply
modelled as follows. The fragmentation and spreading that pan-editing of RNA
entails, intertwines subsequences of $a$ and $b$ modelled by the sequence $%
a_{1}b_{1}a_{2}b_{2}...a_{k}b_{k}$. We assume the DNA and hence this
sequence to be circular, so that $b_{k}$ connects with $a_{1}$, and we call $%
k-1$ the level of fragmentation. Initially, length of subsequence $a_{i}$ is
$n_{a}/k$ and length of subsequence $b_{i}$ is $n_{b}/k$. We model
replication rounds, in which for every round, with probability $p$, deletion
of DNA with random length $l$, at a random position, may occur. Any deletion
in an '$a$' sequence is lethal. Surviving individuals replicate such, that
the smaller their total DNA, the faster they replicate. This advantage
(given as an increased representation in the next round) is modelled by a
function $r(x)$. For example, a complete deletion of $35\%$ ($n_{b}$) of the
DNA will translate in a $3/2$ stronger representation in the next generation.

The process is modelled using a Markov chain approach, with simplifying
assumptions described in the appendix. We study the value $t_{\max }(k)$
monitoring the expected number of generations until no individuals are left
with '$b$' of length $n_{b}$. Any individual, missing part of $b$, can not
survive in the `next' environment ($B$) anymore. The increase of $t_{\max
}(k)$ gives us the added protection that additional fragmentation ($k$)
brings. At level $k$ we have $(n_{k}/k+1)k$ many states in our Markov chain
(see appendix). Realistically, $n_{b}$ is set around $8000$ base pairs and $%
k $ can be over $200$, making the Markov chain too big to handle.

By exploiting symmetries and simplifying our model we approximate our Markov
chain, obtaining a manageable model that we can simulate for realistic
values of $n_{b}$ and $k$. An additional feature of this approximation is
parameter $d$, the confidence level, which allows us to smoothly interpolate
between the simplified and the original Markov chain. Varying $d$ suggests
that already for small values we obtain accurate approximations of the
original process. See appendix and Figure 2.

\section{Results obtained by modelling the effects of increasing
mitochondrial gene fragment dispersal}

In a first simulation (see Figure 1B) we decided to compare our highly
simplified genome with and without a fragmented gene ($k=2$) under partially
relaxed selection (when only part of the genome is needed) with concomitant
strong (intraspecific) competition. We simulated a modest (\emph{e.g.} in
the bloodstream much less of the mt genome is needed) situation in which
most ($2/3$) of the mt DNA is still necessary for survival (environment $A$:
'tsetse'). Starting with $10^{10}$ individuals, we look at the number of
generations needed to reduce the amount of individual organisms that have
retained 'ecological flexibility' (\emph{i.e.} they still contain all
genetic information to be able to make the switch to another part of the
life cycle) to $1$. Without any fragmentation: after $108$ generations no
organism has the possibility to switch to the next stage of the life cycle.
With $1$ split: no such individual is left after $156$ generations (values
obtained starting with $108$ individuals: $96$ and $146$ respectively; with $%
10^{12}$ individuals: $119$ and $169$). See the table below:

\bigskip

\begin{equation*}
\begin{tabular}{|c|ccc|}
\hline
&  & Population size &  \\ \hline
$k$ & $10^{8}$ & \multicolumn{1}{|c}{$10^{10}$} & \multicolumn{1}{|c|}{$%
10^{12}$} \\ \hline
1 (no split) & $96$ & $108$ & $119$ \\
2 (one split) & $146$ & $156$ & $169$ \\ \hline
\end{tabular}%
\end{equation*}

\bigskip
\bigskip

The allowed generation time under periods of partially relaxed selective
pressure has been extended by more than $40\%$. We next altered the 'simple'
simulation by using a fixed deletion size for gene fragments (instead of
allowing the full range of sizes of our previous model) while at the same
time making it more informative by varying the amount of gene fragmentation (%
$k$, see Figure 1C). This (full Markov chain) simulation gave a surprising
result: we obtained a direct, \emph{quasi-linear} correlation between the
degree of dispersal/fragmentation ($k$) and the number of generations after
which a complete population loses ecological competence (see Figure 2). In
other words, the function $t_{\max }(k)$ has an almost linear growth rate.

\section{Comparing the model to values observed}

A triatomine bug (the insect vector used by \emph{Trypanosoma cruzi}) will
contain on average $105$ parasites (Kollien and Schaub, 1998); while a bite
with the tsetse fly will transmit $0-40,000$ (mean, $\sim \emph{3},000$)
\emph{T. brucei} infective metacyclics, giving rise to over $10^{10}$
parasites in a Vervet monkey (\emph{Chlorocebus pygerythrus}) at peak
infection ((Thuita \emph{et al.}, 2008) and references therein). We chose
populations of $10^{8}$, $10^{10}$ and $10^{12}$ parasites in our simple
modelling (see the table) and $10^{10}$ for our full Markov chain approach.
Under the parameter values of our model the size of the population does not
seem of major importance (cf the table).

We chose to model $\sim 65\%$ of the total mt DNA in use, and $\sim 35\%$
free from functional selection (for trypanosomatids in mammalian hosts, much
less mt DNA seems essential). Doubling time in mammals can be $4.5$ hours if
unimpeded by immune responses, while infections can be sustained for months.
Thus, our model should look at effects over many generations. In our basic
approach we compare only two instances ('no split' and 'only one split'),
but allow all possible deletion sizes. In the full Markov chain approach we
use an approximation (depending on $d$) allowing us to model random
deletions of varying sizes.

Because we have modelled optimal dissemination (always strictly
intermingling constitutively used and conditionally used regions), we
tentatively infer that the fragmentation values of Figure 2 are relevant
below $150$. Recall that the \emph{complete} \emph{T. brucei} required set
of gRNAs for \emph{all} lifecycles is about $150$ (Hong and Simpson, 2003).

Surprisingly, only $77$ rapid passages (at least $600$ generations) in mice
of tsetse fly infective \emph{T. brucei} already gave rise to a homogeneous
population of parasites that could no longer infect the insect and did not
display mt (oligomycin sensitive) ATPase activity anymore (Hajduk and
Vickerman, 1981), demonstrating rapid takeover by mt DNA mutants. These
mutants in the end indeed \emph{divided more rapidly} than the wildtype in
blood but did not contain large scale mt deletions. This could of course be
explained by the fact that certain parts of the present-day highly dispersed
mt genome are still necessary and thus protect against large scale deletions
under these circumstances, as described by our model of selection for
linkage.

Takeover by deletion mutants could thus have been even faster, as mt
replication times are possibly limiting when doubling every $4.5$ hrs. This
also means (cf. Figure 2) that at a fragmentation level of between $\sim 10$
and $20$, deletion mutants would contribute about as much to depletion of
the wt population as all other inactivating mutations ($>600$ generations).
For \emph{T. brucei} we would expect a fragmentation level of $>20$ and $%
<<150$, which seems realistic. However, all \emph{T. brucei} parameter
values are very difficult to estimate in real life, because:

\begin{enumerate}
\item It is not known precisely how little mt DNA is essential in the
mammalian host,

\item Population sizes and generation time estimates are complicated by
waves of parasitemia, reflecting the immune response to alternating variant
surface glycoprotein expression. Very limited mt function and repeating
population bottlenecks could explain why \emph{T. brucei} has the most
extensive editing observed (see below).
\end{enumerate}

A further conclusion, shedding light on another point of contention
(Landweber, 2007; Speijer, 2008): physical linkage of all mt genes is not
essential for the model to work. This aspect is important when considering
the cox1 gene in the parasitic diplonemid \emph{Diplonema papillatum}
apparently split up in $\sim 250$-bp fragments, located on individual
unlinked DNA circles (Marande \emph{et al.}, 2005) and the existence of
minicircles with only one gRNA in certain trypanosomatids (though often
physically linked).

As the modelling makes clear, it is only reducing the chance of large
deletions giving replication advantages which is essential. Losing a
substantial fraction of -one gRNA encoding- minicircles (a 'large' deletion)
will still compromise viability directly, regardless of \emph{physical}
linkage. Networks themselves did probably evolve to make takeover by
(deletion) mutants even less likely (Borst, 1991).

\section{Evolution of RNA editing: gene fragment dispersal to counter gene
loss?}

We model fragment dissemination as advantageous to parasites (Speijer, 2006;
Speijer, 2007), but others defend `neutral' models (such as 'Constructive
Neutral Evolution' (Covello and Gray, 1993; Gray \emph{et al.}, 2010)); see
(Lukes \emph{et al.}, 2009; Speijer, 2010; Lukes \emph{et al.}, 2011;
Flegontov \emph{et al.}, 2011; Speijer, 2011). Parasitism is not essential
for our model to function: any free-living ancestor in periodically changing
environments adapting its mt function could favour mt gene fragmentation
when encountering strong intraspecific competition.

Quickly changing environments could also be at the basis of the glycosome
(Hannaert \emph{et al.}, 2003; Gualdron-Lopez \emph{et al.}, 2012). Rapidly
changing oxygen levels would lead to either aerobiosis with mt respiration
or anaerobiosis with glycosomal activity for energy generation in
non-parasitic ancestors. For these ancestors, the lack of selective pressure
on the mt genome has been invoked as the raison d'etre for RNA editing
before (Cavalier-Smith, 1997).

The fact that novel kinetoplastids and diplonemids have been found in anoxic
deep-sea basins recently, make such an ancestor more likely (Lara \emph{et
al.}, 2009; Edgcomb \emph{et al.}, 2011). Indeed, despite corrections to the
kinetoplastid phylogenetic tree (Moreira \emph{et al.}, 2004; Katz \emph{et
al.}, 2012), \emph{pan-editing is still seen as ancestral}, originating 500
to 700 million years ago (Lukes \emph{et al.}, 1994; Fernandes \emph{et al.}%
, 1993). Kinetoplastids are part of the Euglenozoa (with diplonemids and
euglenoids). Interestingly, another form of mt gene fragmentation occurs in
the \emph{diplonemid D. papillatum} (Marande \emph{et al.}, 2005; Marande
and Burger, 2007), as described above.

Transcript editing has become less extensive in more recently evolved
species (\emph{e.g.} compare cox3 editing in \emph{L. tarentolae} and \emph{%
C. fasciculata} with its pan-editing in \emph{T. brucei}). Editing loss
results from reverse transcription of (almost completely) edited RNA
followed by homologous recombination of the cDNA with mt DNA. The 5' and 3'
homology requirements of the cDNA would sometimes result in a need for 5'
editing of the 'new' transcript encoded by the mt DNA upon recombination.
This is actually seen in cox3 and cytb in \emph{L. tarentolae} and \emph{C.
fasciculata} (Landweber, 1992).

Another nice example of recent editing domain length reduction during
cryptogene evolution is found in the ND8 gene of three related insect
trypanosomatids, \emph{again strongly suggestive of correlations between
life cycle complexity and editing extent} (see below; Gerasimov \emph{et al.}%
, 2012). Pan-editing entails a lot of extra complexity and energetic costs,
so losing it is indeed what one would indeed expect, as soon as reduction of
life cycle complexity allows it.

\section{Pan-editing counters gene loss}

In 1993 Covello and Gray (Covello and Gray, 1993) introduced a general model
for the evolution of different RNA editing forms. RNA editing activity is
first acquired by (a combination of) pre-existing enzymes. Mutations at
'editable' nucleotide positions in the genome occur next. Later on, editing
becomes essential after fixation by a chance process in which an altered
form replaces the original without a selective advantage. In the case of
kinetoplastid editing this model could in principle explain the emergence of
a few 'limited' editing instances, but it is very hard to envisage how it
explains the rapid acquisition of multiple instances of pan-editing with
hundreds of gRNAs.

This model does not identify selective pressure(s) responsible for an \emph{%
active} increase of editing potential. This pressure was postulated to
reside in the fact that gene scattering could protect against loss of
temporarily non expressed mt genes during periods of intense intraspecific
competition (Speijer, 2006), by warding off large deletion mutants
outcompeting wildtype kinetoplastids (see above and Figure 1A). Thus,
ecological flexibility is retained, allowing kinetoplastids to occupy highly
diverse (parasitic) niches and undergo extensive speciation (cf. the
repeated evolution of \emph{T. brucei} strains that can not infect their
insect vectors anymore (Lun \emph{et al.}, 2010)).

Pan-editing seems an effective way of making large deletions improbable. It
necessitates presence of \emph{all} cognate small gRNAs (containing
information for the U sequence) to express a cryptogene (encoding the 'GAC
sequence'). These `gene fragments' colonize mt DNA so that every random
large scale deletion will now contain some information still under selective
pressure (Speijer, 2006; Speijer, 2007). The following observations fit our
model.

\begin{enumerate}
\item Especially transcripts encoding gene products which are crucial in all
life cycle stages should be prone to pan-editing (such as ATP6 and RPS12,
encoding components of the \emph{T. brucei} mt F1Fo ATPase and ribosome
respectively). It is exactly because they are always required that
fragmenting their genes with concomitant spread of cognate gRNA genes in mt
DNA leads to efficient 'mt DNA integrity checkpoints'.

\item The role of gRNAs as such mt DNA integrity checkpoints can also be
performed by evenly distributed tRNA genes. Compare \emph{e.g.} tRNA genes
in human mt DNA (Anderson \emph{et al.}, 1981) and \emph{E. coli}'s multiple
rRNA operons distributed in such fashion (Blattner \emph{et al.}, 1997).
However, exceptionally, all trypanosomatid mt tRNA genes are absent, mt
tRNAs coming from the cytoplasm (Hancock and Hajduk, 1990).

\item There seems to be a clear correlation between life cycle complexity
and amount of pan-editing, \emph{i.e.} gene fragment dissemination (see
above and \emph{e.g.} (Gerasimov \emph{et al.}, 2012). \emph{T. brucei},
with its highly complex life cycle, still encodes a COX III cryptogene, much
reduced in length (see above), so its loss in the mammalian host would be
less advantageous. Much more importantly, its dispersed gRNA segments
protect parts (encoding NADH dehydrogenase subunits) of the genome not used
in the fly (Speijer, 2006; Speijer, 2007). Thus, gene fragmentation implicit
in mt RNA editing can be seen as an instance of a general tendency to
'evenly' distribute genetic information over the genome. In this way large
scale deletion of pieces released from selection becomes impossible, no
longer threatening ecological flexibility.

\item Our modelling demonstrates that the number of generations a population
retains individuals with a full mt genome complement increases
quasi-linearly with increases in RNA editing (\emph{i.e.} with its
concomittant gene scrambling). It thus allows \emph{positive selection of
small, incremental increases of editing potential} under the appropriate
circumstances (\emph{e.g.} less frequent environment-host exchange).
\end{enumerate}

Constructive Neutral Evolution, ignoring 'evolvability' aspects tries to
explain pan-editing as resulting from genetic drift and population
bottlenecks only, but our model explains both its rapid spread and current
patterns. These patterns correlate not only with a history of population
bottlenecks, but \emph{specifically with alternating selection pressures on
mt DNA} (Speijer, 2006; Gray \emph{et al.}, 2010; Lukes \emph{et al.}, 2011;
Flegontov \emph{et al.}, 2011; Speijer, 2011).

\section{Kinetoplastid mitochondrial genome evolution}

Every time kinetoplastids readjust to different environments, founder
effects can occur. The observed rapid speciation and development of
unexpected, 'weird' biochemical properties could then be a natural result.
Such founder effects might explain acquisition of (limited) RNA mt
editing/gene fragmentation in Euglenozoa (Walker, 2007) in the first place.
The kind of selection described here would then quickly give pan-editing in
lineages with alternating mitochondrial demands.

Present-day kinetoplastids show an enormous diversity in mt DNA: maxicircles
of up to 200 kb, gRNA encoding minicircles of 1-3 kb with only one to
several gRNA genes, and 200 kb circles possibly encoding hundreds (Simpson
\emph{et al.}, 2000; Maslov and Simpson, 1994; Blom \emph{et al.}, 2000).
Minicircles clearly evolved later, possibly in response to coding capacity
demands of pan-editing. This increased mt DNA size and led to networks
combating minicircle loss (Borst, 1991). As mentioned, over time editing
becomes less extensive in some kinetoplastids (Landweber, 1992;Gerasimov
\emph{et al.}, 2012), reflecting the complexity and overall burden of the
elaborate editing system.

\section{Conclusion}

We have modelled mt gene scrambling, matching it with observations regarding
the frequency of pan-editing and the ecology and population sizes of
trypanosomatids in which it occurs. Important modelling parameters include:
the population size, the deletion size (as a percentage of total genome
size), the probability of a deletion in each new generation and the
replication advantage value for a deletion. The outcome for every simulation
was the number of generations after which no viable individual was left. We
took the \emph{T. brucei} lifecycle to deduce fixed values, only varying
population size and amount of gene fragmentation. Population size did not
seem a major determinant. Our output correlated extent of
fragmentation/dispersal and increase in number of generations retaining
ecological competence, which turned out to be a quasi-linear function. The
'fragmentation value' of \emph{T. brucei} mt DNA seems to be in the range
suppressing large scale deletion population takeover. To prevent such
takeover, condemning parasites to die with their host, a gene architecture
of tremendous complexity evolved. It links survival of genes released from
selection pressure to survival of genes still under such control. Thus,
while fragmentation reduces efficiency it seems to enhance long term
ecological success. We are currently looking at gene fragmentation of \emph{%
Plasmodium falciparum} (Feagin \emph{et al.}, 2012) and ciliates (Nowacki
\emph{et al.}, 2011) in light of this mechanism, with only the malaria
parasite possibly fitting the mould.

\bigskip

\emph{Acknowledgements. }We thank Nick Dekker for his expertise in
preparing Figure 1, Ronald de Wolf for helpful discussions and Andrew Moore
for valuable comments and suggestions to improve the manuscript. This
research has been partially funded by a Dutch BSIK/BRICKS grant and by Vici
grant 639-023-302 from the Netherlands Organization for Scientific Research
(NWO). SS is supported by the Royal Society. Part of this work was performed
while SS was visiting the CWI; its financial support is gratefully
acknowledged.

\newpage

\appendix

\section{Model}

We want to mathematically model the life cycle of genetic material of a
parasite. We need to model the following parameters:

\begin{enumerate}
\item Population size ($s_{p}$);

\item Deletion size;

\item Probability of deletion ($p$);

\item Level of fragmentation/editing ($k$);

\item Number of generations in each host;

\item Replication advantage of smaller DNA ($r(x)$);

\item Stop criterion: number of parasites that still have their full DNA and
will be able to survive in a different host.
\end{enumerate}

The model will enable us to determine the relationship between some of these
parameters.

We will first model the process of replication and deletion in its most
general form and then simplify it to enable us to study/simulate the model
more accurately . We assume there are two environments, in which the
parasite lives. We call these $A$ and $B$. $A$ could be when the parasite,
for example a trypanosome, occupies a tsetse fly and $B$ when it is in the
human host. The level of fragmentation of the mitochondrial DNA of an
individual parasite is modeled as follows. Each parasite has a sequence $a$
base pairs of length $n_a$ necessary to survive in environment $A$ and $b$
of length $n_b$ needed in environment $B$. The total length of it's
mitochondrial DNA will therefore be $n=n_a+n_b$ base pairs. For the
remaining we assume that the parasite is in environment $A$.

The process of editing induces that the base-pairs of $a$ and $b$ are
intertwined modeled as follows by the sequence $a_1b_1a_2b_2\ldots a_kb_k$.
We assume that the DNA and hence this sequence is circular, so that $b_k$
connects with $a_1$, and we call $k$ the amount of fragmentation. Moreover
we assume initially that the length of each subsequence is the same: $a_i$
is $n_a/k$ and $b_i$ is $n_b/k$.

\subsection{Deletion}

During each replication cycle deletion of DNA may occur. We will model this
by setting $p<1$ the probability that a deletion occurs. When a deletion
occurs we model this by picking uniformly at random a point $1\leq $
position $\leq n$ on the cyclic DNA and cut out a uniformly random chosen
length $1\leq l\leq n$. Two things may now happen. Either the resulting
smaller circle of DNA misses part of its DNA that is essential for its
current environment-- for example in environment $A$ all DNA in the parts $%
a_{i}$ are needed and any deletion in such part will kill the parasite-- or
the smaller DNA still has all the $a_{i}$ parts intact and the deletion only
removed a part of the $B$- type DNA. The resulting DNA of a surviving
individual is the sequence $a_{1}^{\prime }b_{1}^{\prime }\ldots
a_{k}^{\prime }b_{k}^{\prime }$ where for each $a_{i}^{\prime
}=a_{i}=n_{a}/k $ and $b_{i}^{\prime }\leq b_{i}$; note that $b_{i}^{\prime
} $ may become $0$.

\subsection{Replication}

Next, the surviving individuals replicate in such a way that the smaller
their total DNA the faster they replicate, which translates into more
descendants (replication advantage). We use a function $r(x)$ that models
the replication advantage. Let $s(x)=\sum_{i=1}^{k}b_{i}$ be the total size
of the $B$-part (number of $B$ base-pairs) of sequence $x=a_{1}b_{1}\dots
a_{k}b_{k}$. Let $\max_{r}$ be the maximum multiplicative replication
advantage, individuals who lost all their $B$-DNA replicate with a factor $%
\max_{r}$ (\emph{e.g.} $3$), and set $\min_{r}$ to be the minimum
replication factor, for individuals who still have their whole DNA intact (%
\emph{e.g.} 2).
\begin{equation}
r(x)=\max_{r}-s(x)\ast (\max_{r}-\min_{r})/n_{b}
\label{eq:multiplication_factor}
\end{equation}%
Note that the function $r(x)$ linearly interpolates between $\max_{r}$ and $%
\min_{r}$, depending on the size of the DNA.

\subsection{Simulation}

We will need to model both the process of deletion and that of replication.
We start by describing the first.

\subsubsection{Deletion}

We use a Markov chain to model our process, though the entries will not be
probabilities since they can be larger than 1. A Markov chain is a graph $%
G(E,N)$, where $N$ is the set of nodes corresponding to all the possible
states of the parasite being alive: the sequences $a_1b^{\prime }_1\ldots
a_kb^{\prime }_k$, with $b^{\prime }_i\leq b_i$. There is a directed edge
from node $x$ to node $y$ in $G$ when sequence $y$ can be obtained from
sequence $x$ via a \emph{single} deletion. The edges of $G$ are labeled with
the probability that these deletions occur. There are also edges (self
loops) from each node $x$ to itself, with label $(1-p)$. These self loops
represent the probability that no deletion occurs.

We next construct a matrix $D$ out of our Markov chain $G(E,N)$ of size $|N|
* |N|$ as follows. The entry $D(x,y)$, is equal to the label of edge $(x,y)$%
, and $0$ when $(x,y)$ is not an edge in $G$. The starting state of our
process corresponds to all the individuals that have all their DNA still
present. This corresponds to the unit vector $v_0$, with $v_0(1)=1$ and $%
v_0(i)=0$ for $2<i\leq|N|$. In other words we fix the first entry of our
vector to correspond to the state where all the DNA is still present.
Likewise entry $x$ corresponds to the fraction of individuals in state $x$.

\subsubsection{Replication}

In order to model the replication process we define the diagonal matrix $%
R(x,x) = r(x)$. Multiplication with $R$ corresponds to replicating state $x$
with replication factor $r(x)$. A single deletion step followed by a
replication step is now simply the matrix $M=RD$.

The vector $v^{\prime }=Mv_{0}$ corresponds to our population after a
deletion and replication step of our process. Note that the vector $%
v^{\prime }$ does not have $L_{1}$ norm $1$ anymore\footnote{%
The $L_{1}$ norm of $v$, $|v|_{1}=\sum_{i=1}^{|N|}|v(i)|$.}. We now need to
take into account the boundary conditions induced by the maximum population
size of the parasites as follows. We would like to view $v_{0}$ and $%
v^{\prime }$ as the probability distributions over the state space.
Initially all the probability mass is on the full DNA state and
progressively this mass flows to other states. We can then interpret the
multiplication of $v_{t}(i)$ with the size of the maximum population $s_{p}$%
, $s_{p}v_{t}(i)$ as the \emph{expected} number of individuals that have DNA
corresponding to state $i$ after $t$ generations. This means that we have to
renormalize our vector: $v_{1}=v^{\prime }/|v^{\prime }|_{1}$ in order to
make it a probability. This completes one full generation step of our
process, and in general:
\begin{equation*}
v_{t}=\frac{Mv_{t-1}}{|Mv_{t-1}|_{1}}
\end{equation*}%
Since $M$ is a linear map, we may renormalize at the end, so that:
\begin{equation*}
v_{t}=\frac{M^{t}v_{0}}{|M^{t}v_{0}|_{1}}
\end{equation*}

Note that implicit in $M$ is the value of $k$, the fragmentation level,
which we have omitted in our notation so far for simplicity. Note that when $%
k$ grows so does the size of the Markov chain. Keeping track of this
parameter we define for each $k$:
\begin{equation*}
v_{t,k} = \frac{M_k^tv_{0,k}}{|M_k^tv_{0,k}|_1}
\end{equation*}
Let $t_{max}(k)$ be the maximum $t$ such that $s_pv_{t,k}(1) \geq 1$. The
value $t_{max}(k) + 1$ tells us the expected number of generations until
there are no individuals left that have their full DNA, at a fragmentation
level $k$. We are interested in the growth rate of this function $t_{max}(k)$%
. 

\section{Reduction of the State Space}

It is easy to see that at fragmentation level $k$ we have $(n_{b}/k+1)^{k}$
many states in our Markov chain. Typically the total size of the
mitochondrial DNA of type $B$ will be around $10,000$ base pairs and $k$ can
be over $200$. This means that for these values the Markov chain becomes way
too big to handle. We will first exploit some symmetries in our problem and
then show how we can approximate the Markov chain in order to obtain a more
manageable formulation that we can simulate.

\subsection{Symmetries}

Since we are only interested in evaluating the first entry of $v_{t,k}$, we
do not need the information of the states $a_1b^{\prime }_1\ldots
a_kb^{\prime }_k$ with $b^{\prime }_i\leq b_i$, we only need to keep track
of how \emph{\ many} blocks we have of size $0 \leq s \leq n_b/k$. For
example the starting state corresponds to the following tuple with $n_b/k+1$
entries $(k,0,\ldots,0$), the first entry indicating how many blocks we have
of size $n_b/k$, the second how many of size one less and the last how many
of size $0$. Following this convention, the fully depleted state becomes $%
(0,\ldots,k)$. So these are precisely the states $(c_1,\ldots,c_{n_n/k +1})$
such that $\sum_{i=1}^{n_b/k+1} c_i = k$. This corresponds exactly to the
number of multi-sets of cardinality $k$ with elements taken from a finite
set of cardinality $n_b/k +1$. This multi-set coefficient is equal to ${%
\binom{(n_b/k) + k }{k}}$ and can be bound from below by $((n_b/k+k)/k)^k$.
This second representation is significantly smaller than our initial set-up,
but still too large for the range of parameters we are interested in. We
therefore have to simplify our process further.

\subsection{Simplification}

\label{sec:simplification} We modeled a deletion of DNA, by randomly picking
a position $pos$ in the (circular) DNA and then remove a piece starting at $%
pos$ of random length. Individuals survived this deletion whenever only DNA
from within a B-block $i$ was deleted. We now simplify this as follows. Fix
a parameter $d$, $1\leq d<n_{b}$, which we will call the \emph{confidence
level}. We will only keep track for each block $i$ whether it has size $%
a\ast n_{b}/(k\ast d)$, with $0\leq a\leq d$. Whenever a random deletion
left us with a block size
\begin{equation*}
a\ast n_{b}/(k\ast d)\leq b_{i}^{\prime }<(a+1)\ast n_{b}/(k\ast d),
\end{equation*}%
we set the block size $b_{i}^{\prime }=a\ast n_{b}/(k\ast d)$, while keeping
the probability of this event the same~\footnote{%
Strictly speaking we should write $\lceil a*n_{b}/(k*d)\rceil $, we
approximate this and assume that $n_{b}$ is divisible by $d*k$.}. We thus
give slightly more probability to deleting larger parts within block $i$. We
will see later that this change is not very significant.

For example, setting $d=1$, models that whenever a deletion falls within
block $i$, we completely remove block $i$ (\emph{i.e.} it will have size $0$%
). On the other hand for $d=n_{b}/k$ we get back our old process. Confidence
level $d$ thus allows us to interpolate smoothly between the simplified
process and the original process.

For confidence level $d$, the states of our Markov chain will be $d+1$%
-tuples $(c_1,\ldots,c_{d+1})$ such that $\sum_{i=1}^{d+1} c_i = k$. The
first entry indicates the number of blocks that have size $n_b/k=(n_b/k*d)*d$%
, the second entry describes the number of blocks that have size $%
(n_b/k*d)*(d-1)$, and the last entry the number of blocks that have size $0
= (n_b/k*d)*0$.

The number of states we have in our Markov chain, with fragmentation $k$ at
confidence level $d$ is equal to the number of multi-sets of cardinality $k$
out of a finite set of cardinality $d+1$, which is ${\binom{k+d }{k}}$.

\subsection{Transition Probabilities}

Here we will make precise the transition probabilities between any pair of
states in our Markov chain. Given any state $x=(c_{1},\ldots ,c_{d+1})$ such
that $\sum_{i=1}^{d+1}c_{i}=k$. The transition from $x$ to $x$, \emph{i.e.}
no deletion occurred, is labeled with $(1-p)\ast r(x)$, where $r(x)$ is
taken as in equation~\ref{eq:multiplication_factor} with $s(x)$ the size
function adapted for these simplified states:
\begin{equation*}
s(x)=n_{a}+\sum_{i=0}^{d}c_{i+1}\frac{n_{b}(d-i)}{k\ast d}
\end{equation*}%
Transition from $x=(c_{1},\ldots ,c_{d+1})$ to $x^{\prime }=(c_{1}^{\prime
},\ldots ,c_{d+1}^{\prime })$ are only possible if there is an $i<j$ such
that $c_{i}=c_{i}^{\prime }+1$ and $c_{j}=c_{j}^{\prime }-1$, and for all
the other indices $i^{\prime }$ the states are the same: $c_{i^{\prime
}}=c_{i^{\prime }}^{\prime }$. This guarantees that exactly one $B$-block of
size corresponding to $i:n_{b}(d+1-i)/(k\ast d)$ transforms, by means of a
deletion, to a block of size corresponding to $j:n_{b}(d+1-j)/(k\ast d)$.
The probability that this transition occurs turns out to be:
\begin{equation}
\frac{c_{i}\ast m(j,k,d)}{s(x)^{2}}\ast p  \label{eq:probability}
\end{equation}%
Where $m(j,k,d)$ is the number of ways one can transform a block of length
corresponding to $i$, i.e of length $n_{b}\ast (d+1-i)/(k\ast d)$, to a
block of length corresponding to $j$, using the rule of rounding down
described in section~\ref{sec:simplification}. Note that this number only
depends on $j,k$, and $d$ and not on $i$. For example if $i=1$ and $j=2$
this corresponds to the number of ways one can delete a sequence of length $%
1 $ up-to $n_{b}/(k\ast d)$ in a sequence of length $n_{b}/k$, which is
equal to
\begin{equation*}
n_{b}/k+(n_{b}/k-1)+\ldots +(n_{b}/k-n_{b}/(k\ast d)+1).
\end{equation*}%
In general this becomes
\begin{equation*}
m(j,k,d)=\sum_{i^{\prime }=\frac{n_{b}\ast (d-(j-1))}{k\ast d}+1}^{\frac{%
n_{b}\ast (d+1-(j-1))}{k\ast d}}i^{\prime }
\end{equation*}

In equation~\ref{eq:probability} we divide by $s(x)^2$ because each possible
deletion has probability $s(x)^2$ to occur at a fixed position and is of a
fixed length. Finally we multiply in equation~\ref{eq:probability} with $p$,
the probability that a deletion occurs.

\section{Results}

We started by simulating our process with a confidence level of $1$. This
gives a state space of size $k+1$ with the corresponding simulation matrix
of size $(k+1)^{2}$. We ran the simulation with the following parameters: $%
p=10^{-6}$, total size of the mitochondrial DNA is $26\ast (10^{3})$, with $%
34\%$ part being of type B. We set the maximum population size $%
s_{p}=10^{10} $. The replication advantage was computed as in equation~\ref%
{eq:multiplication_factor} with $\max_{r}=3$ and $\min_{r}=2$. The function $%
t_{\max }(k)$ when plotted for values of $k$ ranging from $1$ to $200$ shows
a nearly perfect line, indicating that the advantage of fragmentation is
almost linear. We were able to show rigorously that in the simple case of $%
d=1$ the fragmentation advantage can never be more than linear, that is, we
were able to show a linear upper bound on the function $t_{\max }(k)$. This
was done by studying the spectrum of a simplified $2\times 2$ Markov chain.

Next we simulated the same process with increasing confidence values $d$.
These results show that in each case, for these parameters, we get almost
straight lines each one with a slightly steeper slope. However for
successive values of $d$ the \emph{increase} of the slope appears to be
halving each time. This suggests that already for a small value of $d$ we
have a reasonably good approximation of our original Markov chain.

\begin{figure}[!b]
  \begin{center}
    \includegraphics[width=3.5in]{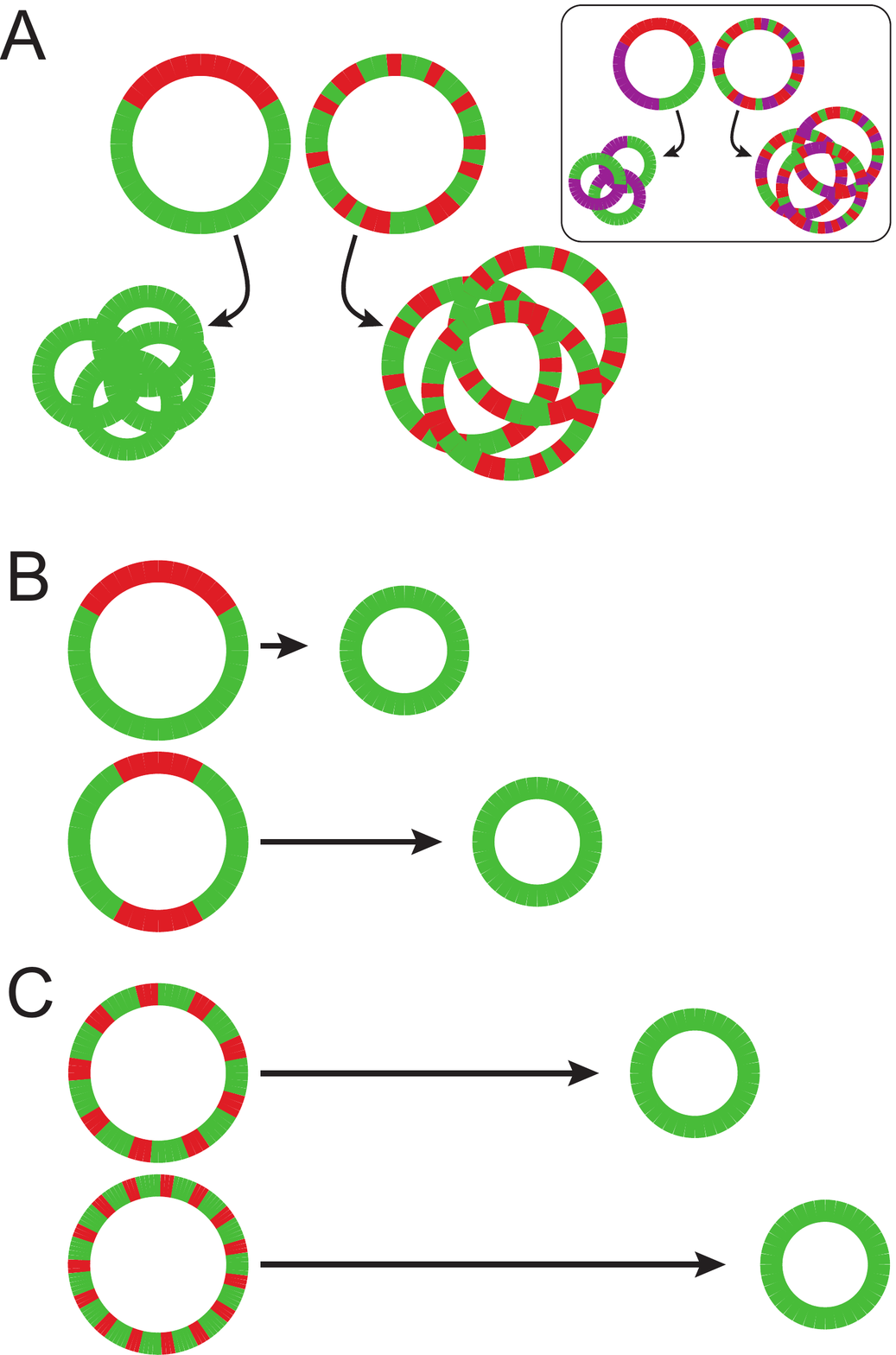}
  \end{center}

  \caption{\small Schematic representation of gene scrambling in organisms alternating between
life cycles with different selective pressures and strong intraspecific
competition (A) and of modelling used in this study (B,C). Inset:
kinetoplastids are represented by multicoloured mt DNA circles with 2 or 3
classes of genes: 'green' genes required in one life cycle (\emph{e.g.}
mammalian blood for \emph{T. brucei}), 'red' genes required in another (%
\emph{e.g.} Tsetse fly) and 'violet' genes required in both. Only
presence/absence of selective pressure is important in modelling. Full
figure: red (absence of selective pressure) and green only. Arrow lengths
(not to scale) represent number of generations upon which the population
loses ecological competence ($t_{\max }(t)$ of our model). (A) Red genes
disappear during many generations of strong competition. Gene scrambling
will slow down this process. (B) Simulations of instances of scrambling over
successive generations. In the first simulation we deduce how many
generations more the population is allowed by a split before losing
ecological flexibility (represented by the smaller green circle), using many
possible deletions. (C) In the second simulation we deduce how many
generations more retain full flexibility by stepwise scrambling increase,
using complete segment deletions only.}
  \label{fig-label}
\end{figure}

\begin{figure}[!b]
  \begin{center}
    \includegraphics[width=6in]{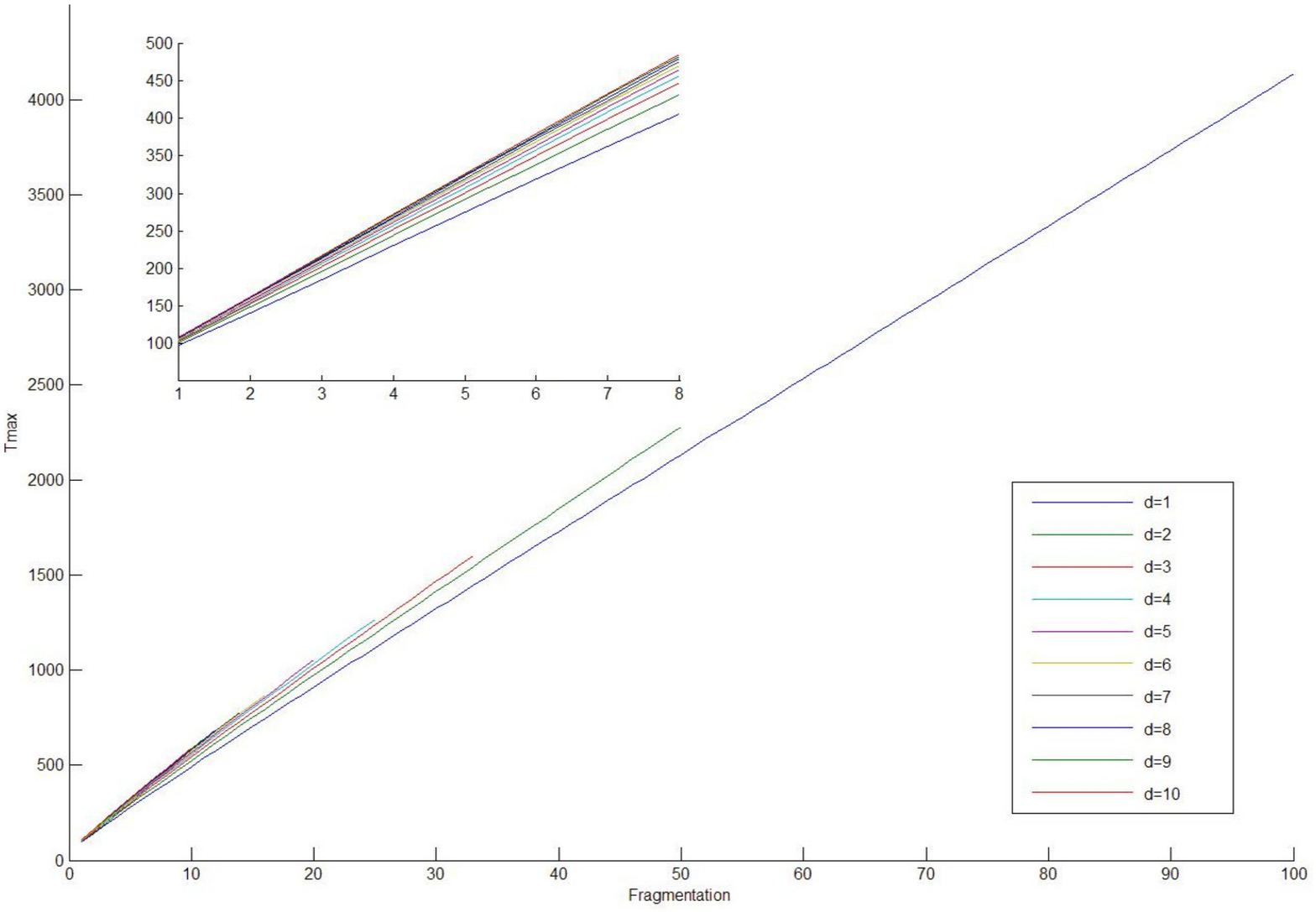}
  \end{center}

  \caption{\small Quasi-linear correlation between amount of fragments ($x$-axis, parameter $k$
of our model) and maximum number of generations upon which the population
loses ecological competence ($y$-axis; $t_{\max }(k)$ of our model). Inset:
close-up of fragmentation levels $1-8$; inset on the right: colour code for
different confidence levels ($d$). The higher $d$, the better the
approximation (see appendix).}
  \label{fig-label}
\end{figure}


%
%
%
%
%

\end{document}